\documentstyle[11pt,twoside,jltp]{article}

\title{Short Time Quantum AC Response of a System of Nanomagnets}

\author{G. Rose$^{1}$ and P. C. E. Stamp$^{1,2}$}
\address{$^{1}$ Department of Physics and Astronomy, and Canadian
Institute for Advanced Research,
University of British Columbia, Vancouver B.C. V6T 1Z1 Canada\\
$\;\;$ \\
$^{2}$ Institut Laue-Langevin, Ave. des Martyrs, Grenoble 38042, France \\
}

\runninghead{G. Rose and P.C.E. Stamp}{Short Time AC response of a
system of Nanomagnets}

\begin{document}

\begin{abstract}
We calculate the magnetization relaxation 
in the short-time regime for an ensemble of 
nanomagnets in the presence of a low frequency external
AC biasing field at temperatures lower than the magnetic
anisotropy energy of the individual nanomagnets. 
It is found that the relaxation is strongly affected by AC fields with amplitude
larger than that of the $T_2$ fluctuations in the nuclear field.
This will allow experimental probing of the nuclear spin relaxation mechanism.

PACS numbers:
\end{abstract}

\maketitle

\section{INTRODUCTION}
Recent experiments\cite{exp1,exp2,exp3,exp4,exp5,exp6,exp7} on ensembles of magnetic
macromolecules have shown evidence for resonant tunneling relaxation. Two of these
experiments show evidence of tunneling in the ``quantum regime'', where only the
two lowest levels of each molecule are occupied; in the ``Mn-12'' molecule\cite{exp6}, this
happens below a crossover temperature $T_c \approx 2 K$ and in the ``Fe-8'' molecule\cite{exp7},
below $T_c \approx 0.4 K$. Experiments in the Fe-8 system have gone down to 70 $mK$, with no
change in the relaxation characteristics below 0.36 $K$; this constitutes {\it prima facie} evidence for
a quantum regime in Fe-8 molecular crystals\cite{PS1,add1}. Theoretical work\cite{PS1,PS2,add2} on
the ``quantum relaxation'' below $T_c$ indicates that intermolecular dipole coupling and hyperfine
coupling to the nuclear spins are necessary to explain the relaxation characteristics below $T_c$.
A number of predictions based on this theory have recently been verified experimentally\cite{add2,add3};
this work is discussed elsewhere in this volume\cite{add4}. In particular, the prediction of a universal
short-time ``square root'' relaxation has been verified, with a 
characteristic time $\tau_Q$
which depends on both dipolar and hyperfine interactions (as well as the tunneling matrix element
$\Delta_0$).

At present there is no direct measure of $\Delta_0$ in these systems, making it hard to verify whether the
observed $\tau_Q$ is that predicted by theory. In this paper we give 
preliminary results of a theory of quantum relaxation of ensembles of 
nanomagnets 
in the presence of an applied AC field. We show that the relaxation characteristics are strongly
altered, in a way which should allow (i) the determination of $\Delta$, and (ii) the demonstration that
$\tau_Q$ is controlled by nuclear spins as well as by dipolar interactions.

We begin by considering an ensemble of nanomagnets or magnetic macromolecules. The high energy 
Hamiltonian for such systems has been shown to flow under reduction of temperature to a fixed point
effective Hamiltonian given by 
\begin{equation}
H=\sum_{\vec{r}} H^{(0)}(\vec{\tau}, \{ \vec{\sigma}_k \} ) + \sum_{\vec{r},\vec{r}^{'}} H_D 
	(\vec{\tau}_z^{\vec{r}}, \vec{\tau}_z^{\vec{r}^{'}}) + \sum_{k,k^{'}} V (\vec{\sigma}_k,
		\vec{\sigma}_{k^{'}})
\end{equation}
where $\vec{\tau}^{\vec{r}}$ is a Pauli matrix acting at 
molecular site $\vec{r}$, in the Hilbert
space of the two lowest molecular 
states, and $\vec{\sigma}_k$ is a Pauli matrix acting on the two
relevant states of the $k^{th}$ nuclear spin; we assume that $k=1..N$. The internuclear
term $V(\vec{\sigma}_k,\vec{\sigma}_{k^{'}})$ is usually dipolar, and $|V_{k,k^{'}}| \approx
1-100 kHz$ (and measurable as $T_2^{-1}$, in for example nuclear spin echo experiments). Below $T_c$,
$H_D$ is diagonal in $\vec{\tau}_z$ (molecular flip-flop transitions are rare); it causes a bias
$\xi_D(\vec{r})=\sum_{\vec{r}} V_D(\vec{r}-\vec{r}^{'}) \vec{\tau}_z^{\vec{r}^{'}}$ at site $\vec{r}$,
which varies over a scale $E_D \approx 0.5 K$ around the sample. In the absence of nuclear spins
\begin{equation}
H^{(0)}( \vec{\tau}^{\vec{r}} ) = \Delta_0 \vec{\tau}_x^{\vec{r}}
\end{equation}
but in general one has\cite{TUP}
\begin{center}{$
H^{(0)}(\vec{\tau}^{\vec{r}}, \{ \vec{\sigma}_k \} ) = 
\Delta_0 [\cos \{ \Phi +\sum_{k=1}^{N} \alpha_k \vec{n}_k \bullet \vec{\sigma}_k
		\} \hat{\tau}_{+} +H.C.] $}\end{center}
\begin{equation} + {{\hat{\tau}_{z}}\over{2}}
		\sum_{k=1}^{N} \omega_k^{||} \vec{l}_{k} \bullet \vec{\sigma}_k
		+{{1}\over{2}} \sum_{k=1}^{N} \omega_k^{\perp} \vec{m}_k
		\bullet \vec{\sigma}_k
\end{equation}

This complex fixed point Hamiltonian contains all coupling effects between the molecular spin
at $\vec{r}$ and the surrounding nuclear spins\cite{PS1,TUP}. Of crucial importance are the
diagonal hyperfine couplings $\omega_k^{||}$ to each $\vec{\sigma}_k$ (varying between $\approx
0.1 mK-0.5 K$ in different systems), and the complex dimensionless amplitude $\alpha_k$ for
$\vec{\sigma}_k$ to flip when $\vec{\tau}^{\vec{r}}$ does. $\vec{l}_k, \vec{m}_k$ and $\vec{n}_k$ 
are unit vectors, and $\Phi$ is a renormalized Kramers-Berry-Haldane phase 
(given by $\Phi = \pi S$ when ${{1}\over{2}} \sum_{k=1}
^N |\alpha_k|^2 << 1$).

Typically, experiments begin by first polarizing the system, with all molecules aligned, and
then watching the magnetization $M(t)$ decay with time $t$. Here we assume zero applied field,
but add an AC field $H_{ac}(t) =\sum_{\vec{r}} A \cos(\omega t) \vec{\tau}_z^{\vec{r}}$. The
question that we wish to now ask is this--in the experimentally relevant region 
${{1}\over{2}} \sum_{k=1}
^N |\alpha_k|^2 << 1$, ie. where no nuclear spins flip during tunneling, 
what will the magnetization of the crystal look like at short times in the 
presence of this AC field?

\section{THE GENERALIZED MASTER EQUATION}

We may write the magnetization of our system in the form
\begin{equation}
M(t)=\sum_{\vec{r}} \int d\xi M({\vec{r}},\xi,t) = \sum_{\vec{r}} \int d\xi 
	(2 P_{\uparrow}({\vec{r}},\xi,t) -1 )
\end{equation}
where $P_{\uparrow}({\vec{r}},\xi,t)$ is the normalized probability of the 
central spin at
site $\vec{r}$ to be 
``up'' (ie. in state $\vert S_z = +S \rangle$) and in a static bias $\xi$
at time $t$. 

A solution for $P_{\uparrow}({\vec{r}},\xi,t)$ over timescales $\approx
O(1/\omega)$ is
rather messy. However experimentally one is usually interested in
relaxation over much
longer timescales, in which case one can write a kinetic or ``master''
equation of
the form
\begin{center}{$
{\dot{P}_{\alpha}
({\vec{r}},\xi,t)=-W(A,\omega;\xi) \{P_{\alpha}(\vec{r},\xi,t)
  - P_{-\alpha}(\vec{r},\xi,t) \}}$}\end{center}
\begin{equation}
-\sum_{\vec{r}^{'},\alpha^{'}} \int d\xi ^{'} W(A,\omega,\xi^{'})
 [ P_{\alpha \alpha ^{'}}^{(2)}(\vec{r}, \vec{r}^{'}; \xi,\xi ^{'} ;t) - P_{\alpha \alpha ^{'}}^{(2)}
(\vec{r},\vec{r}^{'};\xi-\alpha \alpha ^{'} V_D(\vec{r}-\vec{r}^{'}), \xi ^{'};t) ] 
\end{equation}
where $P^{(2)}$ is the usual 2-molecule distribution function. Note that
we have assumed that nuclear $T_2$ fluctuations have decorrelated each
pass of the local bias field through resonance from the others; this
implies that $\omega < NT_2^{-1}$.
When $A$ or $\omega$ are zero, the master equation reduces to a ``static'' kinetic 
equation\cite{add1}, in which $W(A,\omega;\xi) \rightarrow \tau_N^{-1} (\xi) 
\sim
(\Delta^2/ \xi_0) e^{- \vert \xi \vert / \xi_0}$,
the nuclear spin driven transition rate\cite{PS1,add1}. We ignore higher order multimolecular
terms $P^{(3)}, P^{(4)}$, etc. and assume approximate factorization of $P^{(2)}$. As before\cite{add1}, this
means that the results we derive are only valid when $1-M(t)/M_0<<1$, where $M_0$ is the saturation magnetization.

Our method generalizes that given for a static applied field; we first solve for the rate function $
W(A,\omega;\xi)$ for a single molecule\cite{PS1}, but now in an AC field, so that the total 
longitudinal field acting on $\vec{\tau}_x^{\vec{r}}$ is
\begin{equation}
\xi_{tot}^{\vec{r}}=\xi(\vec{r}) + A \cos(\omega t) + \delta \xi_{\vec{r}}(t)
\end{equation}
in which $\delta \xi_{\vec{r}}(t)$ is the rapidly varying component coming from $T_2$ fluctuations,
and $\xi(\vec{r})=\xi_D(\vec{r})+\xi_N(\vec{r})$ is the slowly varying sum of dipolar fields
and longitudinal hyperfine fields ($\xi_N(\vec{r}) = \omega_0 M_{\vec{r}}$ for a system with a single
hyperfine coupling $\omega_0$ and total nuclear polarization $M_{\vec{r}}=\sum_{k=1}^{N} < \vec{\sigma}_k^z >$
along the molecular easy axis).

In what follows we assume (as noted above) that no nuclear spins flip during tunneling--the general results
including nuclear flips will be published elsewhere. We also assume that the nuclear spins are in a thermal
ensemble with $kT>>\omega_0$ (it is easy to show that the AC field will drive them into such a high-T
distribution; and all experiments so far have $kT>>\omega_0$ anyway).

In present experiments on magnetic macromolecules where the molecular spin $S \approx O(10)$, an
AC field of amplitude $\delta H = A/g \mu_B S$ equal to 1 $G$ is equivalent to a bias amplitude
$A \approx 50 MHz \approx 25 mK$. Experiments can range roughly 
between $10^{-5} G < \delta H < 100 G$ (
depending on $\omega$). Thus we will assume that $\Delta \xi > A >> \Delta$, where $\Delta
\xi \approx 0.5-1 K$ is the total spread in $\xi$ caused by dipolar and hyperfine fields;
however the ratio $A/\xi_0$ is arbitrary. We then find the following results.

(i)  $A/\xi_0 >>1$ (large AC amplitude). The transition rate $W(A,\omega;\xi) \approx
W_0(\xi/A) \Theta(A^2-\xi^2)$, where
\begin{equation}
W_0(\xi/A) =  
	{{\Delta^2}\over{\sqrt{A^2-\xi^2}}}
\end{equation}
for $ {A-|\xi| >> \xi_0^2/2A}$ and
\begin{equation}
W_0(\xi/A)=  {{\Delta^2}\over{\xi_0}}
\end{equation}
for $ {A-|\xi| < \xi_0^2/2A}$.

(ii) $A/\xi_0 <1$ (small AC amplitude). In this regime the dynamics are controlled
by the nuclear $T_2$ field; one gets, for the transition rate,
\begin{equation}
W(A,\omega;\xi) \approx {{\Delta^2}\over{\xi_0}} e^{-|\xi|/\xi_0}
\end{equation}

The essential effect of the AC field, when $A>>\xi_0$, is to spread the resonant tunneling
over a much larger energy range $2A$ (but at a reduced rate, except when $|\xi| \approx A$,
and the AC field itself varies slowly, so that $\dot{\xi}(t)$ is dominated by $T_2$
fluctuations). As the amplitude of the AC field is decreased the dynamics become
completely dominated by nuclear $T_2$ effects with the crossover occuring at $A \approx \xi_0$.

\section{SHORT TIME DYNAMICS}

The solution for $M(t)$ is now straightforward, following our previous 
methods\cite{add1}.
For large AC amplitude one gets
\begin{equation}
\dot{M}(t)=-{{2 M(t)}\over{\pi}}  \int_{-A}^{A} d\xi {{\Gamma(t)}\over
		{[\xi-E(t)]^2+\Gamma^2(t)}} W_0 (\xi / A)
\end{equation}
for an ellipsoidal sample, where the internal field has Lorentzian spread  
$\Gamma(t)={{4 \pi^2 E_D}\over{3^{5/2}}} (1-M(t))$
and mean $E(t)={{c E_D}} (1-M(t))$, and c depends on the specific geometry of our ellipsoid.
For small AC amplitude one gets the square root relaxation found previously\cite{add1}.

The solution to the large AC amplitude problem (10) is clearly not square root at short
times; in fact one gets
\begin{equation}
1-\bar{M}(t) \approx e^{-t/\tau_{ac}} \hspace{1in} 
( \,\, 1-\bar{M}(t) << A/E_D \,\,)
\end{equation}
where $\bar{M}(t) = M(t)/M_0$; the rate $\tau_{ac}^{-1} \approx \Delta^2 / A$, which
is a factor $E_D/A$ faster than the relaxation ``rate'' $\tau_Q^{-1} \approx \Delta^2/E_D$
which one obtains for an ellipsoid in the square root regime\cite{add1}. 
At longer times, once the internal
field $E(t)>>A$, we get
\begin{equation}
1-\bar{M}(t) \approx \sqrt{t/\tau_Q} \hspace{1in} 
(\,\, 1-\bar{M}(t) >> A/E_D \,\,)
\end{equation}
Clearly if $A/E_D$ is not too small, we may never see a clear square root relaxation;
correlations in $P^{(2)}$ will emerge before the square root does.

For a non-ellipsoidal sample we find in general that the exponential relaxation 
rate will become 
$(\tau_{ac}^{inhom})^{-1} \approx  \Delta^2/E_D \equiv (A/E_D)
\tau_{ac}^{-1}$, since the internal fields are now spread over a large range $\approx E_D$ in
bias space. In all but specially-shaped samples, this will be what is observed.

It is clear from these results that AC experiments can test the nuclear spin mediated
tunneling mechanism, since the prediction here is that it is only when $A>\xi_0$ that
significant deviations from the static square root relaxation will appear. 
This prediction is quite different from what one would find if 
we assumed that tunneling proceeded at a rate $\Delta$ for molecules near exact resonance
($|\xi|< \Delta$); in this case the square root relaxation would break down for $A>\Delta$.
Since in the Mn-12 and Fe-8 systems, $\Delta \approx 10^{-10}-10^{-9} K$, this would imply
a breakdown of the square root law once $A$ exceeded roughly $10^{-8}-10^{-7} G$,
a very small value indeed!

Notice also that since $\tau_{ac}^{-1} \approx  \Delta^2/A$, we have 
a way of determining the important parameter $\Delta$ directly in experiments, 
knowing $A$. It is clear that the results of AC experiments conducted near
$H = 0$ in magnetic macromolecular crystals will give a crucial test of 
present theory.

\section*{ACKNOWLEDGEMENTS}
This research is supported by NSERC, the CIAR, and the ILL; PCES thanks Prof. P. Nozieres for his hospitality in Grenoble.


\begin{thebibliography}{9}

\bibitem{exp1} C. Paulsen and J.G. Park, in ``{\it Quantum Tunneling of Magnetization-
		QTM'94}'' (ed. L. Gunther and B. Barbara), Kluwer publishing, pp. 189-
		207 (1995).
\bibitem{exp2} M. Novak and R. Sessoli, pp. 171-188, {\it ibid}.
\bibitem{exp3} D.D. Awschalom {\em et.al.}, Science {\bf 258}, 414 (1992) and refs.
		therein.
\bibitem{exp4} B. Barbara {\em et.al.}, J. Mag. Magn. Mat. {\bf 140-144}, 1825 (1995).
\bibitem{exp5} J.R. Friedman {\em et.al.}, Phys. Rev. Lett., {\bf 76}, 3830-3833 (1996).
\bibitem{exp6} L. Thomas {\em et.al.}, Nature {\bf 383}, 145-147 (1996).
\bibitem{exp7} C. Sangregorio {\em et.al.}, Phys. Rev. Lett., {\bf 78}, 4645 (1997). 
\bibitem{PS1} N.V. Prokof'ev and P.C.E. Stamp, J. Low Temp. Phys. {\bf 104}, 143 (1996).
\bibitem{PS2} P.C.E. Stamp, pp. 101-197 in ``{\it Quantum Tunneling in Complex Systems}'', 
ed. S. Tomsovic (World Scientific, 1998).
\bibitem{TUP} I. Tupitsyn, N.V. Prokof'ev, P.C.E. Stamp, Int. J. Mod. Phys. {\bf B11},
		2901 (1997).
\bibitem{add1} N.V. Prokof'ev and P.C.E. Stamp, Phys. Rev. Lett. (in press); cond-mat
	9710246.
\bibitem{add2} T. Ohm, C. Paulsen, preprint (Dec. 20, 1997); and see this volume.
\bibitem{add3} L. Thomas, B. Barbara, preprint (Jan 16, 1998); and see this volume.
\bibitem{add4} N.V. Prokof'ev and P.C.E. Stamp, this volume.
\end{thebibliography}
\end{document}